\documentclass[aip,pof,preprint,amsmath,amssymb,floatfix]{revtex4-2}
\usepackage{graphicx}
\usepackage{bm}
\usepackage{booktabs}
\usepackage{siunitx}
\usepackage{xcolor}
\usepackage{microtype}
\usepackage{mathtools}
\usepackage{array}
\graphicspath{{./}}
\begin{document}

\title{Gaussian kinetic representations of rarefied nonequilibrium flows}

\author{Ehsan Roohi}
\email{eroohi@umass.edu}
\affiliation{Department of Mechanical and Industrial Engineering, University of Massachusetts Amherst, Amherst, Massachusetts 01003, USA}

\date{\today}

\begin{abstract}
Compact representations of rarefied flows must preserve kinetic observables, not only smooth macroscopic fields.  We introduce Gaussian kinetic representations for discrete velocity method (DVM)-Shakhov solutions of normal shocks and a lid-driven cavity.  A positive log-density phase-space model reconstructs shock velocity distribution functions (VDFs) and their moments, while a moment-field model compresses wall-bounded cavity structure.  Log-density training recovers heat flux, stress, and third- and fourth-order shock moments without explicit moment supervision; the cavity representation gives a compact continuous wall-transport map.
\end{abstract}

\maketitle

Rarefied gas data are difficult to reduce because the quantities that define the physics are not limited to density ($\rho$), velocity ($\bm u$), and temperature ($T$).  Heat flux, stress, skewness, and fourth-order content are carried by the velocity distribution function (VDF), so a reduced object may look accurate in low-order fields while being wrong in the kinetic observables that control transport and wall loading.  This hierarchy is classical in kinetic theory and direct simulation Monte Carlo (DSMC), where the VDF is the primary unknown.\cite{Bird1994,Cercignani2000,Sone2007,RoohiScanlon2010}  Model Boltzmann equations and discrete velocity methods (DVMs) provide deterministic quadrature-based references for the same observables.\cite{BGK1954,Shakhov1968,Mieussens2000}  Unified gas-kinetic schemes and rarefied-flow reviews further show why rarefied-flow data must be judged through transport-sensitive moments rather than only through $\rho$, $\bm u$, and $T$.\cite{XuHuang2010,RoohiStefanov2016,AkhlaghiRoohiStefanov2023}  Gaussian representations are not new to kinetic theory: Mott-Smith's bimodal shock ansatz, ultra-sparse isotropic-Gaussian approximations of shock-mixed velocity distributions, and recent Gaussian-mixture-model (GMM) compression of distribution functions all use Gaussian components in one form or another.\cite{MottSmith1951,AlekseenkoGrandilliWood2020,HuPlasmaGMM2025}  Separately, localized Gaussian primitives have been used for continuum turbulent-flow representation, where primary variables may be accurate while derivative-sensitive diagnostics degrade.\cite{ShenoyFrankel2026}  The present contribution is therefore not the generic use of Gaussians.  It is the rarefied-flow target: joint space--velocity log-density representation for shocks, full-quadrature recovery of heat flux, stress, and higher moments from the trained $\hat f$, and a wall-bounded moment-field representation that quantifies how compression affects transport observables.  The method is complementary to operator learning and rarefied-flow scientific-machine-learning surrogates: it compresses a converged kinetic state into a continuous representation that can be interrogated off the original storage grid.\cite{LuDeepONet2021,RoohiMahdaviDeepONet2026,RoohiShojaSaniPoF2026,RoohiShojaSaniStefanov2026,RoohiMahdaviNozzle2026,RoohiMahdaviLowRank2026}

\begin{figure*}[t]
\centering
\includegraphics[width=0.82\textwidth]{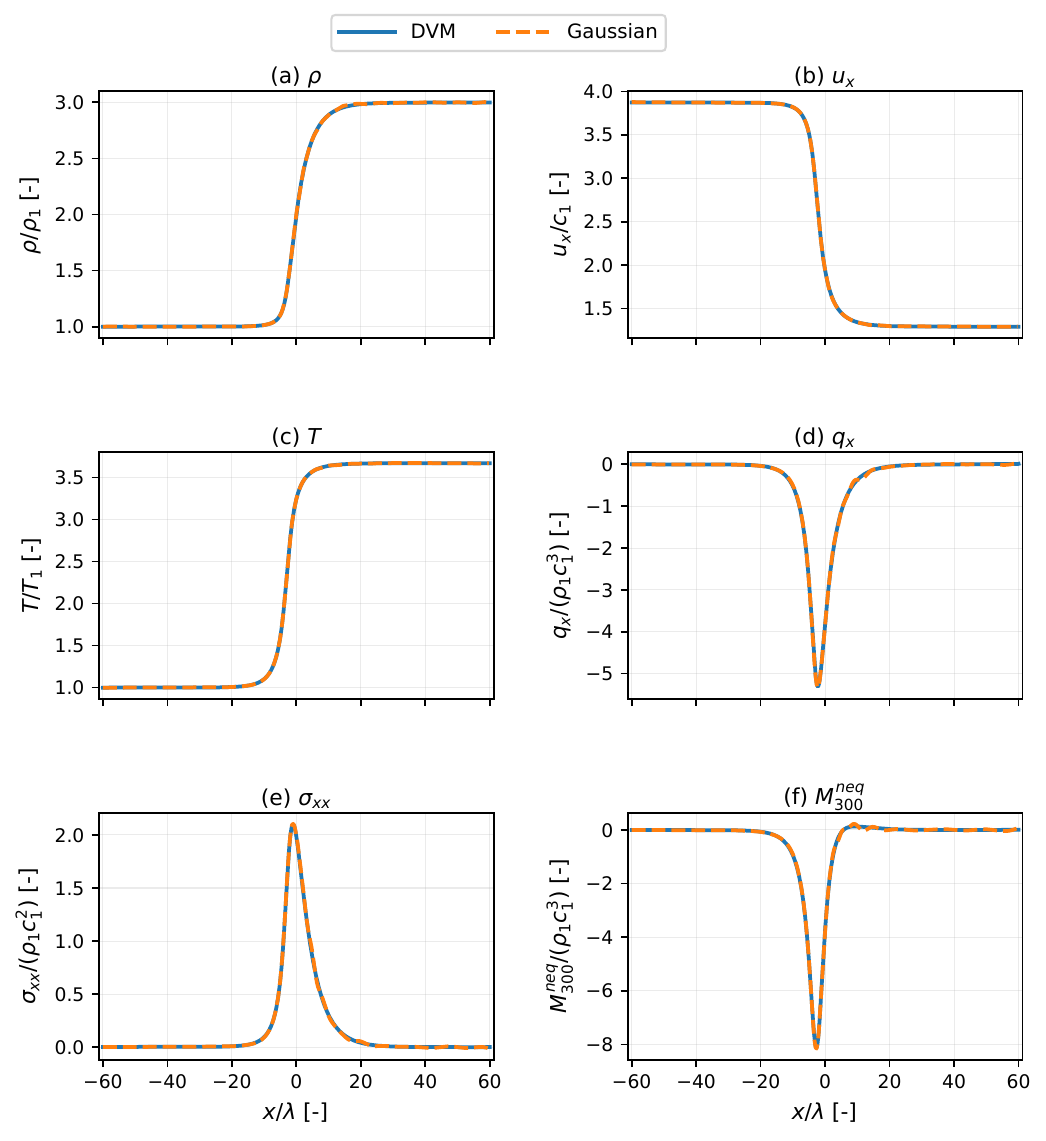}
\caption{M3 normal shock reconstructed by the Gaussian phase-space representation.  All moments are obtained by integrating the same trained $\hat f(x,\xi_x,\xi_y,\xi_z)$.}
\label{fig:m3}
\end{figure*}

\begin{figure*}[t]
\centering
\includegraphics[width=0.82\textwidth]{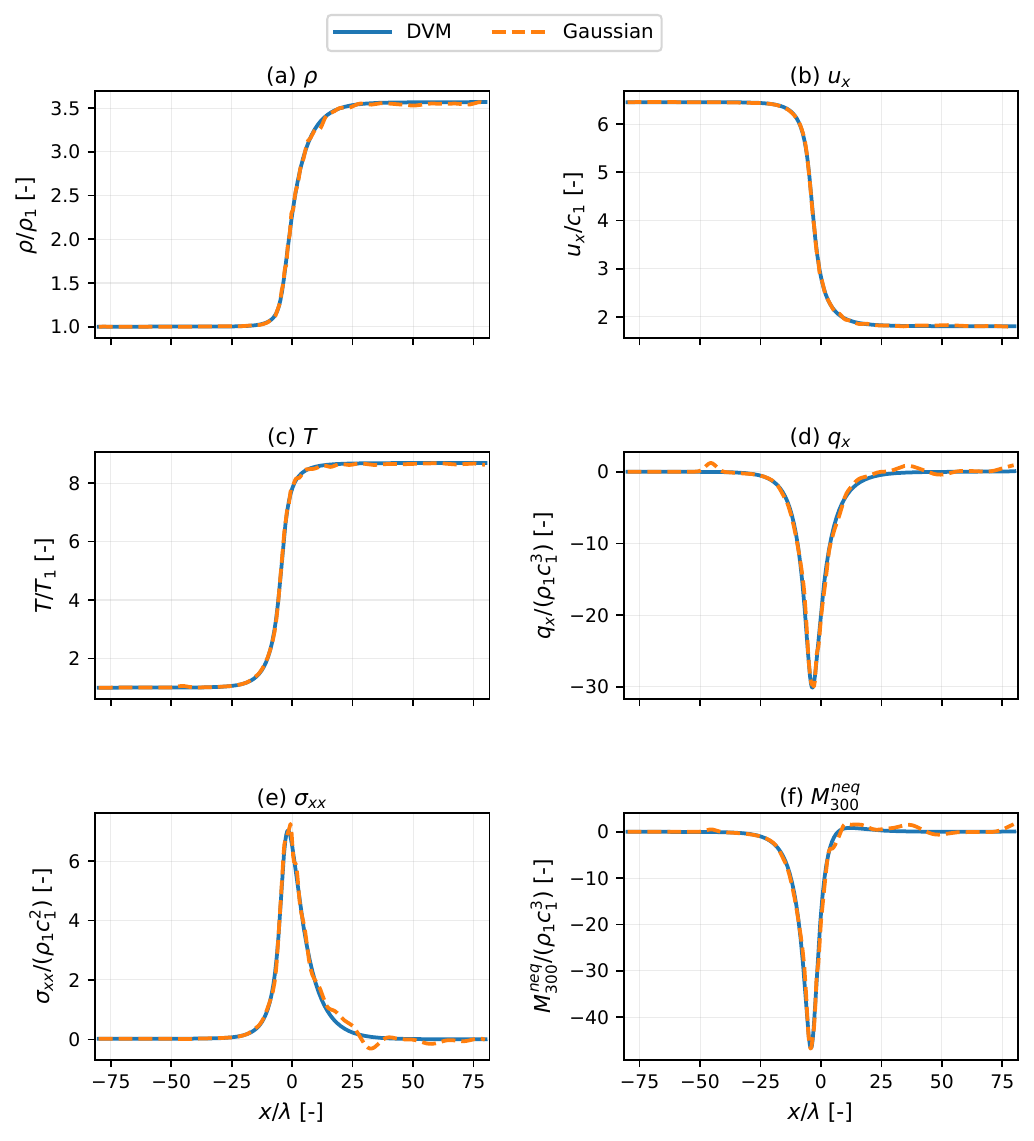}
\caption{M5 normal shock.  Stronger nonequilibrium increases stress and third-order moment error, but the compact phase-space representation preserves the main kinetic structure.}
\label{fig:m5}
\end{figure*}

We use DVM-Shakhov reference solutions for two tests.  The first is a monatomic normal shock at Mach numbers $M=3$ and $M=5$, where $M$ denotes the upstream Mach number and the VDF changes sharply across the shock layer and tail-sensitive moments peak near the shock center.  The second is a lid-driven cavity at Knudsen number $\mathrm{Kn}=0.075$, $U_{lid}=0.1$, and $T_w=1$, where wall heat flux, thermal stress, and shear stress coexist in a two-dimensional geometry and where rarefied-cavity benchmarks are well established.\cite{ZhuRoohiEbrahimi2023,RafieenasabRoohiManela2026}  Shakhov relaxation is used because heat flux is a target observable; a Bhatnagar--Gross--Krook (BGK) reference with Prandtl number $\mathrm{Pr}=1$ would distort one of the moments used here to judge fidelity.  The model equation is
\begin{equation}
\partial_t f+\bm\xi\cdot\nabla_{\bm x}f=\frac{f^S-f}{\tau},
\label{eq:shakhov}
\end{equation}
with
\begin{equation}
 f^S=f^M\left[1+(1-\mathrm{Pr})\frac{\bm c\cdot\bm q}{5pT}\left(\frac{|\bm c|^2}{T}-5\right)\right],\quad
 f^M=\frac{\rho}{(2\pi T)^{3/2}}\exp\left(-\frac{|\bm c|^2}{2T}\right),
\label{eq:shakhov_target}
\end{equation}
where $\bm c=\bm\xi-\bm u$ is peculiar velocity, $p=\rho T$ is pressure, $\bm q$ is heat flux, $\mathrm{Pr}=2/3$ is the Prandtl number, $\tau=\mu(T)/p$ is the relaxation time, and $\mu/\mu_1=(T/T_1)^\omega$ with $\omega=0.81$ is the viscosity law used by the DVM solver.  Shock variables are nondimensionalized by the upstream density, temperature, and thermal speed, denoted by subscript 1; cavity variables are nondimensionalized by length $L$, lid speed $U_{lid}$, wall temperature $T_w$, reference density $\rho_0$, and thermal speed $c_0$.  The shock coordinate in Figs. \ref{fig:m3}--\ref{fig:m5} is $x/\lambda=x^*/\mathrm{Kn}_{eff}$, where $\lambda$ is the upstream mean free path, $x^*$ is the solver coordinate in $[-1/2,1/2]$, $\mathrm{Kn}_{eff}=1/120$ for $M=3$, and $\mathrm{Kn}_{eff}=1/160$ for $M=5$.  With discrete velocities $\{(\bm\xi_n,w_n)\}_{n=1}^{N_v}$, where $N_v$ is the number of velocity nodes and $w_n$ is the quadrature weight, the reference moments are
\begin{align}
\rho &= \sum_n f_n w_n, &
\rho u_i &= \sum_n \xi_{i,n} f_n w_n, \label{eq:macro}\\
3\rho T &= \sum_n |\bm c_n|^2 f_n w_n, &
\sigma_{ij} &= \sum_n c_{i,n}c_{j,n}f_nw_n-\rho T\delta_{ij},\\
q_i &= \frac12\sum_n |\bm c_n|^2c_{i,n}f_nw_n, &
M_{300}^{neq} &= \sum_n c_{x,n}^{3}f_nw_n,\\
M_{400}^{neq} &= \sum_n c_{x,n}^{4}f_nw_n-3\rho T^2 . \label{eq:noneq}
\end{align}
In Eqs. (\ref{eq:macro})--(\ref{eq:noneq}), $i$ and $j$ denote Cartesian components, $f_n=f(\bm\xi_n)$, $\bm c_n=\bm\xi_n-\bm u$, $\delta_{ij}$ is the Kronecker delta, $\sigma_{ij}$ is the stress deviation from isotropic pressure, $q_i$ is heat flux, and $M_{300}^{neq}$ and $M_{400}^{neq}$ are third- and fourth-order deviations in the streamwise peculiar velocity.  The $M=3$ and $M=5$ shock references contain $1800\times64009$ and $2200\times102789$ phase-space samples, respectively, using tensor velocity grids $121\times23\times23$ and $141\times27\times27$ with velocity cutoffs $|\bm\xi|_{max}=16$ and 22, where $|\bm\xi|_{max}$ denotes the maximum retained molecular-speed magnitude in the nondimensional velocity grid.  The cavity reference uses isothermal diffuse walls, a $65\times65$ physical grid, 41 velocity nodes per direction over $[-5,5]$, and a final residual $8.2\times10^{-8}$; the stored state contains 20 DVM-Shakhov moment fields.  These deterministic references avoid DSMC sampling noise and make the representation error itself observable.

For shocks we fit the VDF itself in normalized phase space $\bm z=(\tilde x,\tilde\xi_x,\tilde\xi_y,\tilde\xi_z)$ through a positive log-density Gaussian mixture,
\begin{equation}
\log \hat f_\theta(\bm z)=\log\sum_{m=1}^{N}\exp\left[a_m-\frac12Q_m(\bm z)\right],
\label{eq:phase_gaussian}
\end{equation}
where $N$ is the number of Gaussian kernels, $m$ labels a kernel, $a_m$ is the learned log-amplitude, $Q_m$ is its positive quadratic form, $z_\ell$ is component $\ell$ of $\bm z$, $\mu_{m\ell}$ is the kernel center, and $s_{m\ell}>0$ is the width.  The diagonal form uses $Q_m=\sum_{\ell=1}^{4}[(z_\ell-\mu_{m\ell})/s_{m\ell}]^2$.  A correlated $x$-$\xi_x$ block was also tested,
\begin{equation}
Q_m^{x\xi_x}=\frac{\alpha_m^2-2r_m\alpha_m\beta_m+\beta_m^2}{1-r_m^2},\quad
\alpha_m=\frac{\tilde x-\mu_{mx}}{s_{mx}},\quad
\beta_m=\frac{\tilde\xi_x-\mu_{m\xi_x}}{s_{m\xi_x}},
\label{eq:xvx}
\end{equation}
where $r_m\in(-1,1)$ is a learned correlation coefficient, $\alpha_m$ and $\beta_m$ are normalized offsets in $x$ and $\xi_x$, and diagonal transverse terms are retained for $\xi_y$ and $\xi_z$.  Centers are initialized from mass-weighted samples and all amplitudes, centers, and scales are optimized.  The principal loss is a Huber loss in $\log f$,
\begin{equation}
\mathcal L_f=\left\langle H_\delta\{\log\hat f_\theta(\bm z)-\log [f(\bm z)+\epsilon]\}\right\rangle .
\label{eq:loss_logf}
\end{equation}
Here $\mathcal L_f$ is the training loss, $H_\delta$ is the Huber penalty with threshold $\delta$, $\epsilon$ is the log floor used to avoid evaluating $\log 0$, and $\langle\cdot\rangle$ denotes a minibatch average.  Sampled moment penalties are used only as ablations; reported shock moments are always recomputed from $\hat f_\theta$ by the same full quadrature as the DVM reference.  For the cavity we fit a continuous field map,
\begin{equation}
\hat{\bm U}(x,y)=\bm b+
\sum_{m=1}^{N}W_m(x,y)\bm A_m,
\quad
W_m=\frac{\exp[-(\bm r-\bm\mu_m)^T\bm D_m^{-1}(\bm r-\bm\mu_m)]}{\sum_k\exp[-(\bm r-\bm\mu_k)^T\bm D_k^{-1}(\bm r-\bm\mu_k)]},
\label{eq:moment_field}
\end{equation}
where $\hat{\bm U}$ is the fitted moment vector, $\bm b$ is a bias vector, $W_m$ are normalized Gaussian weights, $\bm r=(x,y)$, $\bm\mu_m$ is the physical-space center, $\bm D_m=\mathrm{diag}(s_{mx}^2,s_{my}^2)$ contains squared widths, $\bm A_m\in\mathbb{R}^{20}$ is the output-amplitude vector, and the denominator sums over all kernels $k$.  The parameter count is $24N+20$.  The state vector is
\begin{multline}
\bm U=(\rho,u,v,T,q_x,q_y,\Theta_x,\Theta_y,\Theta_z,
\sigma_{xx},\sigma_{yy},\sigma_{xy},\\
M_{3x},M_{3y},S_x,S_y,M_{4x},M_{4y},K_x,K_y).
\label{eq:Uvector}
\end{multline}
Here $\Theta_i$ are directional temperatures, $M_{3i}$ and $S_i$ are third-order indicators, and $M_{4i}$ and $K_i$ are fourth-order or closure-sensitive indicators used by the DVM post-processing.  Unlike the shock model, Eq. (\ref{eq:moment_field}) is a moment-field representation rather than a positive VDF, so realizability is not built in; its value is as a compact, continuous diagnostic of wall-transport fields.  In particular, consistency identities among fitted moment channels are not imposed by construction, and this distinguishes the cavity test from the phase-space shock representation.  Training used Adam stochastic optimization with log floors, bounded widths, minibatched phase-space or spatial samples, and the settings listed in the supplementary material; shock errors use full velocity quadrature at the reported evaluation stations and cavity errors use the full $65\times65$ physical grid.

Figure \ref{fig:m3} shows the $M=3$ shock.  With $N=512$ diagonal Gaussians trained only on $\log f$, the model captures both Rankine-Hugoniot-scale variation and localized nonequilibrium extrema.  The relative $L_2$ error, $\|G-D\|_2/\|D\|_2$ with Gaussian prediction $G$ and DVM reference $D$, is $1.42\times10^{-3}$, $6.56\times10^{-4}$, and $6.48\times10^{-4}$ for $\rho$, $u_x$, and $T$, while $q_x$, $\sigma_{xx}$, $M_{300}^{neq}$, and $M_{400}^{neq}$ remain within $1.48\%$, $1.68\%$, $2.03\%$, and $2.07\%$.  Thus, for this case, the moment hierarchy is recovered without explicit heat-flux, stress, or high-moment losses: an accurate log-density approximation carries the relevant nonequilibrium structure.  The nominal shock compression is large: for diagonal phase kernels, the factor $9N$ counts one log-amplitude, four centers, and four widths per kernel, so $N=512$ gives 4608 parameters instead of $1.15\times10^8$ stored $f$ samples for $M=3$.

For the stronger $M=5$ shock, Fig. \ref{fig:m5} shows the expected increase in stress and tail sensitivity.  The plotted case uses the sampled moment-informed loss; low-order errors remain below $0.7\%$, while $q_x$, $\sigma_{xx}$, $M_{300}^{neq}$, and $M_{400}^{neq}$ are in the $7$-$8\%$ range.  A log-density $M=5$ study is deferred to the extended article; the reported $M=5$ case therefore characterizes the moment-informed representation, not the optimal objective at high Mach number.  The available ablation table in the supplementary material nevertheless shows an important point: moment-informed training is not automatically superior.  Noisy sampled moment penalties can degrade stress recovery, and adding kernels does not substitute for physically meaningful placement.

\begin{figure*}[t]
\centering
\includegraphics[width=0.90\textwidth]{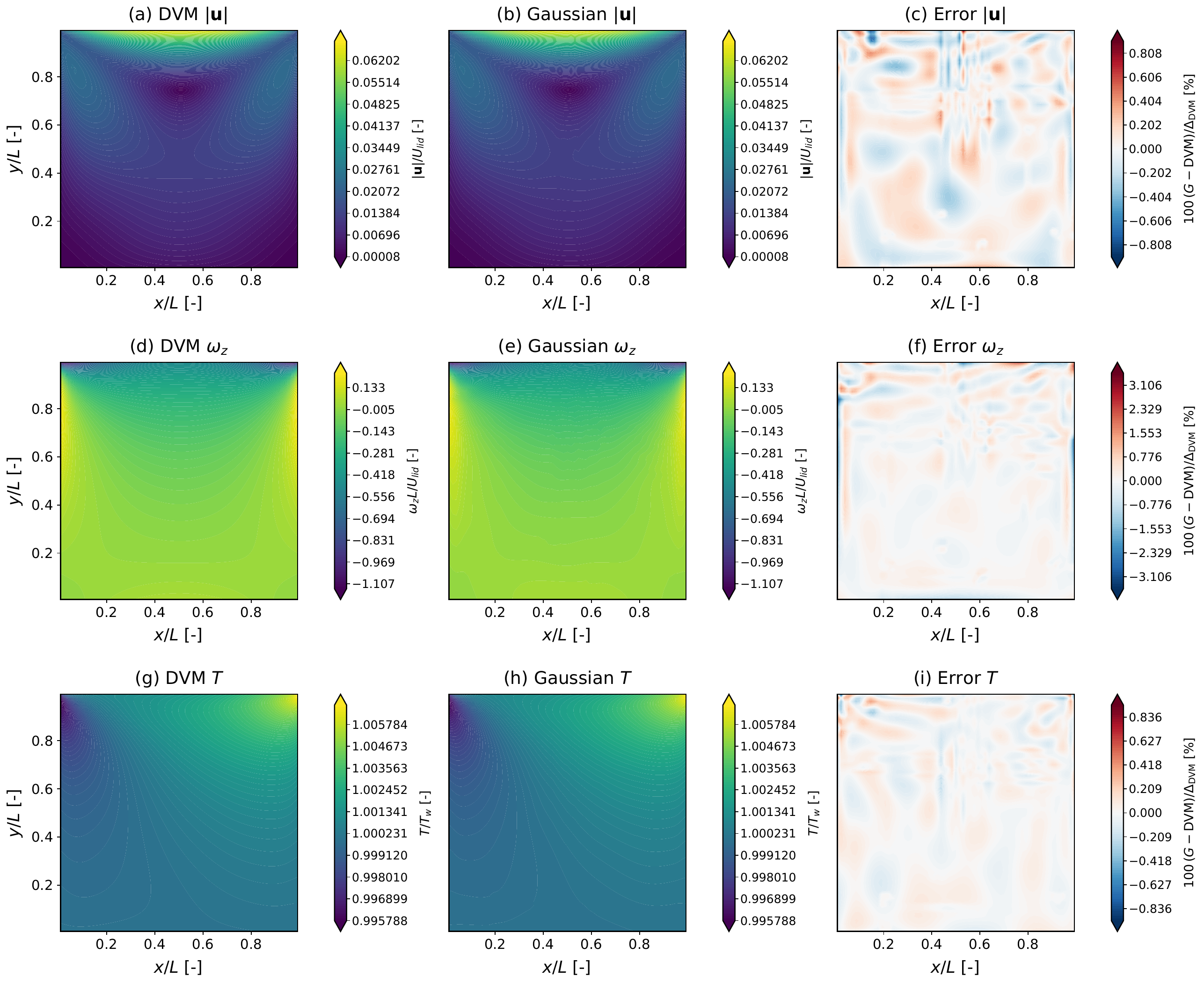}
\caption{Cavity physical fields at $\mathrm{Kn}=0.075$ reconstructed by the $N=256$ Gaussian moment-field representation.  Error maps are percentages of the DVM field range.}
\label{fig:cavity_physical}
\end{figure*}

\begin{figure*}[t]
\centering
\includegraphics[width=0.90\textwidth]{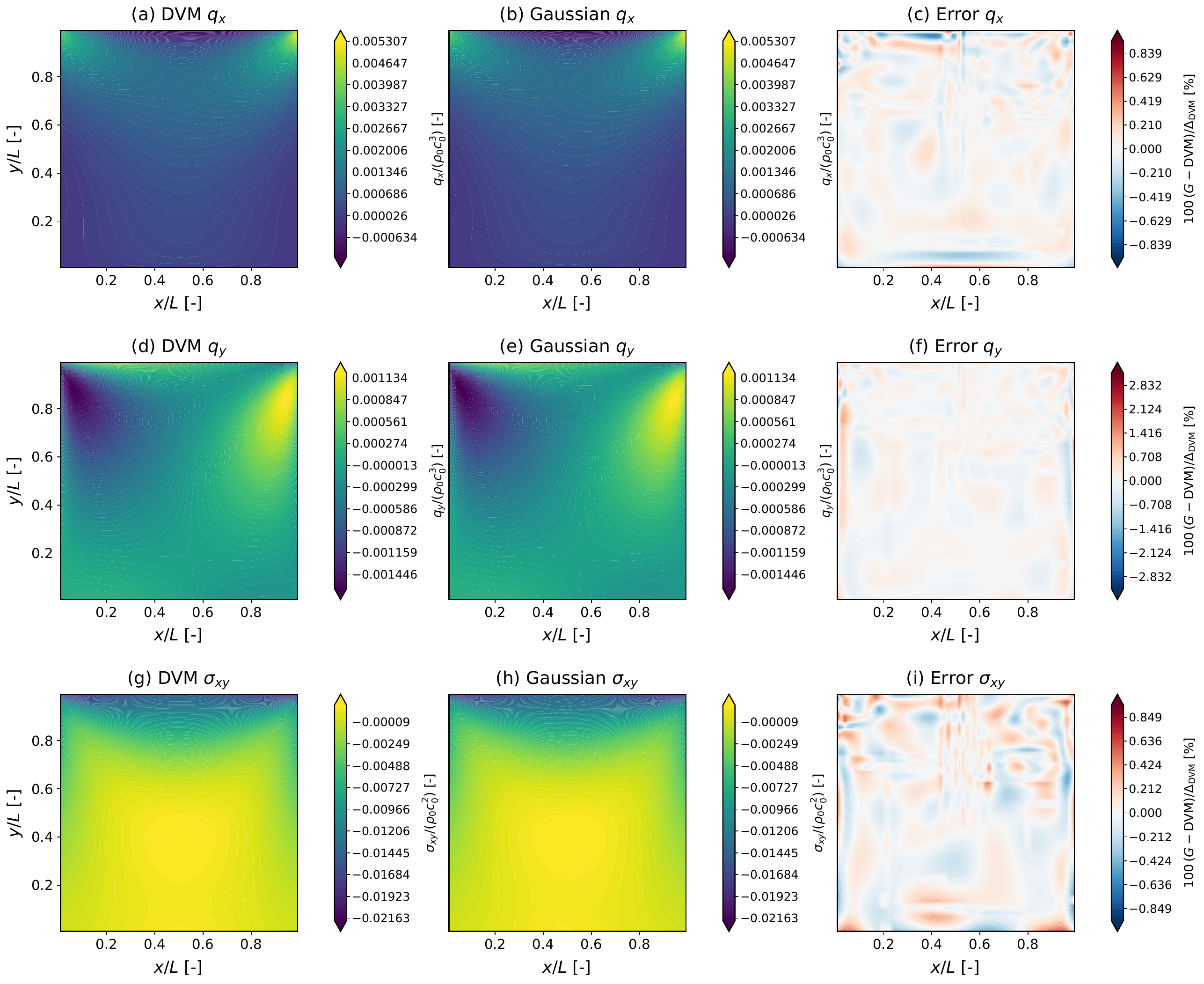}
\caption{Cavity nonequilibrium fields.  Heat-flux and shear-stress structures are recovered by the same continuous representation used for the low-order fields.}
\label{fig:cavity_noneq}
\end{figure*}

The cavity tests whether localized Gaussians retain wall-bounded transport topology rather than simply storing a small array.  Figure \ref{fig:cavity_physical} shows $|\bm u|$, $\omega_z$, and $T$ for $N=256$.  Figure \ref{fig:cavity_noneq} shows $q_x$, $q_y$, and $\sigma_{xy}$, the wall-transport analogues of shock nonequilibrium moments.  Error contours are percentages of the DVM field range, $100(G-D)/\Delta_{D}$, where $G$ is the Gaussian field, $D$ is the DVM field, and $\Delta_D=\max(D)-\min(D)$; this avoids singular pointwise percentages near zeros.  At $N=256$, the selected errors are $0.837\%$, $1.455\%$, and $0.741\%$ for $q_x$, $q_y$, and $\sigma_{xy}$, respectively.  The compression-fidelity sweep in Table \ref{tab:summary} shows that $N=64$ already gives $54\times$ compression with a selected-field maximum error of $4.35\%$, whereas $N=512$ reduces that maximum to $0.326\%$ at $6.9\times$ compression.

\begin{table*}[t]
\caption{Representative compression and relative-error summary.  Shock compression is nominally relative to stored $f$ samples; cavity compression is relative to $65\times65\times20$ stored moment values.  The shock fourth-order entry uses $M_{400}^{neq}=\overline{c_x^4}-3\rho T^2$, not the raw fourth moment.}
\label{tab:summary}
\centering
\scriptsize
\begin{ruledtabular}
\begin{tabular}{lcccccc}
Case & representation & $N$ & comp. & low-order max & $q/\sigma/M_3$ max & $E_{M_4^{neq}}$\\
\hline
M3 shock & phase $\log f$ & 512 & $2.5\times10^4$ & 0.00142 & 0.02030 & 0.02066\\
M5 shock & phase moment & 512 & $4.9\times10^4$ & 0.00663 & 0.07724 & 0.07841\\
Cavity & moment field & 64 & 54.31 & -- & 0.04354 & --\\
Cavity & moment field & 128 & 27.33 & -- & 0.02719 & --\\
Cavity & moment field & 256 & 13.71 & -- & 0.01455 & --\\
Cavity & moment field & 512 & 6.87 & -- & 0.00326 & --\\
\end{tabular}
\end{ruledtabular}
\vspace{2pt}
{\footnotesize Cavity selected maximum is over $U,V,q_x,q_y,\sigma_{xy},M_{3x},M_{3y}$; full 20-channel errors are supplied with the data.}
\end{table*}

These results reposition Gaussian compression for rarefied flows.  The novelty is not the use of Gaussians alone: Mott-Smith shock mixtures, the ultra-sparse isotropic-Gaussian approximation of shock-mixed velocity distributions by Alekseenko, Grandilli, and Wood, and recent plasma GMM compression already show that local velocity distributions can be approximated by Gaussian components.\cite{MottSmith1951,AlekseenkoGrandilliWood2020,HuPlasmaGMM2025}  The present contribution is the coupled kinetic target: a positive phase-space log-density representation for shocks, full-quadrature recovery of heat flux, stress, and third- and fourth-order deviations from the trained $\hat f$, explicit comparison against sampled moment-loss ablations, and a wall-bounded moment-field representation that exposes how compression affects transport observables.  A per-cell velocity-space mixture can approximate local or shock-mixed VDF shape without providing a smooth coupled representation in $x$ and $\bm\xi$; conversely, a generic neural operator can predict fields without revealing where kinetic information is concentrated.  The Gaussian parameters instead define a compact set of locations, scales, and amplitudes that can be interrogated as a kinetic database layer.  This is also why compression should not be judged by one smooth field norm.  A reduced object may reproduce $\rho$ and $T$ while losing the odd, mixed, and fourth-order deviations that control heat transfer, wall loading, and closure sensitivity.  Conversely, once $\log f$ is represented accurately, nonequilibrium moments can be recovered without storing the full DVM phase-space array.  Analytic polynomial moments of Gaussian components are possible over unbounded velocity domains, but the present errors are deliberately reported by the truncated DVM quadrature so that the comparison is exactly aligned with the reference solution.

The method is therefore best viewed as a compact, differentiable, storage-grid-independent layer on top of DVM or DSMC data, not as a replacement for those solvers.  Its centers and widths are geometrically interpretable: they locate where representational capacity is allocated and what support scale is used.  The remaining errors identify where the representation is physically strained--high-Mach shock stress, tail-sensitive third moments, and localized cavity wall transport.  Future work should combine the phase-space and cavity strategies by fitting full wall-bounded VDF data, enforcing conservation and realizability constraints during training, and adapting kernel density according to heat-flux, stress, or entropy-production indicators.  The present tests also caution that sampled moment penalties and larger kernel counts are not automatically beneficial; the training strategy should be selected by the most fragile observable of interest.  See the supplementary material for the full shock ablation, cavity sweep, training settings, and accessible figure descriptions.

\bigskip
\noindent\textbf{Author declarations.} \textbf{Conflict of Interest.} The author has no conflicts to disclose.  \textbf{Author Contributions.} E.R. conceived the study, generated and analyzed the data, implemented the Gaussian representations, and wrote the manuscript.

\noindent\textbf{Data availability.} The DVM reference data, trained Gaussian parameters, and shock and cavity plotting scripts are available from the corresponding author upon reasonable request and will be deposited in a public repository upon acceptance.

\end{document}

\begin{center}
{\Large Supplementary Material for ``Gaussian kinetic representations of rarefied nonequilibrium flows''}
\end{center}

\noindent This supplementary material gives ablation, provenance, and training information referenced in the Letter. No additional simulations beyond those reported in the main text are introduced here. The only added diagnostic is the fourth-order nonequilibrium deviation $M_{400}^{neq}=\overline{c_x^4}-3\rho T^2$, where $c_x$ is the streamwise peculiar velocity, $\rho$ is density, and $T$ is temperature; it is computed from the already shipped predicted-moment files and DVM references.

\section*{S1. Shock ablation table}
\begin{table}[h!]
\centering
\small
\caption{Relative $L_2$ moment errors for shock reconstructions. ``Moment'' denotes sampled moment-informed training; final errors are evaluated by full DVM quadrature. The fourth-order entry is the nonequilibrium deviation, not the raw fourth moment.}
\resizebox{\textwidth}{!}{%
\begin{tabular}{lccccccccc}
\toprule
Case & Model & $N$ & training & $E_\rho$ & $E_{u_x}$ & $E_T$ & $E_{q_x}$ & $E_{\sigma_{xx}}$ & $E_{M_{300}}$ / $E_{M_{400}^{neq}}$\\
\midrule
M3 & diag & 512 & $\log f$ & 0.00142 & 0.00066 & 0.00065 & 0.01475 & 0.01681 & 0.02030 / 0.02066\\
M3 & diag & 512 & moment & 0.00935 & 0.00306 & 0.00514 & 0.07969 & 0.17843 & 0.08235 / --\\
M3 & $x$-$\xi_x$ & 512 & moment & 0.00934 & 0.00300 & 0.00476 & 0.08058 & 0.17266 & 0.08644 / --\\
M3 & $x$-$\xi_x$ & 1024 & moment & 0.00940 & 0.00280 & 0.00541 & 0.10423 & 0.17762 & 0.10515 / --\\
M5 & diag & 512 & moment & 0.00663 & 0.00244 & 0.00506 & 0.07029 & 0.07724 & 0.07632 / 0.07841\\
\bottomrule
\end{tabular}}
\end{table}

\noindent The M5 log-density-only ablation was not run. The main text therefore avoids claiming that the M3 log-density training trend necessarily persists at higher Mach number; the reported M5 result is a transparent assessment of the available moment-informed training case.

\section*{S2. Cavity compression-fidelity sweep}
\begin{table}[h!]
\centering
\small
\caption{Cavity sweep for selected wall-transport fields. Compression is relative to $65\times65\times20$ stored scalar values. The full fitted state has 20 channels; the selected-field maximum shown below is over $U,V,q_x,q_y,\sigma_{xy},M_{3x},M_{3y}$.}
\resizebox{\textwidth}{!}{%
\begin{tabular}{rrrrrrrrrrrr}
\toprule
$N$ & params & comp. & $E_U$ & $E_V$ & $E_{QX}$ & $E_{QY}$ & $E_{\sigma_{xy}}$ & $E_{M3X}$ & $E_{M3Y}$ & max & mean\\
\midrule
64 & 1556 & 54.30591 & 0.01960 & 0.02041 & 0.02173 & 0.04354 & 0.02236 & 0.03894 & 0.02901 & 0.04354 & 0.02794\\
128 & 3092 & 27.32859 & 0.01130 & 0.01225 & 0.01304 & 0.02719 & 0.01535 & 0.01999 & 0.01718 & 0.02719 & 0.01661\\
256 & 6164 & 13.70863 & 0.00624 & 0.00551 & 0.00837 & 0.01455 & 0.00741 & 0.01062 & 0.00879 & 0.01455 & 0.00878\\
512 & 12308 & 6.86545 & 0.00170 & 0.00160 & 0.00322 & 0.00326 & 0.00214 & 0.00263 & 0.00243 & 0.00326 & 0.00243\\
\bottomrule
\end{tabular}}
\end{table}

\section*{S3. DVM provenance and training settings}
Shock variables were nondimensionalized by the upstream state.  The stored shock coordinate is $x/\lambda=x^*/\mathrm{Kn}_{eff}$ with $x^*\in[-1/2,1/2]$, $\mathrm{Kn}_{eff}=1/120$ for $M=3$, and $\mathrm{Kn}_{eff}=1/160$ for $M=5$.  The Shakhov reference used $\mathrm{Pr}=2/3$ and $\mu/\mu_1=(T/T_1)^{0.81}$.  The shock velocity grids were $121\times23\times23$ and $141\times27\times27$ with cutoffs 16 and 22 for $M=3$ and $M=5$, respectively.  The cavity case used isothermal diffuse walls, a $65\times65$ physical grid, 41 velocity nodes per direction over $[-5,5]$, and final residual $8.2\times10^{-8}$.

Shock phase-space models used $N=512$ unless otherwise noted, initialized from $5\times10^5$ sampled phase points. The loss was a Huber loss in $\log f$ with the error clipped at 20, log floor $10^{-35}$, base learning rate $5\times10^{-4}$, center and width learning-rate multipliers of 0.5 and 0.25, minibatches of 24 spatial locations and 384 velocity samples per location, and 80,000 optimization steps. For sampled moment-informed runs, the moment loss used 1024 velocity samples, 12 spatial samples, and weight 0.05 every five steps. The cavity moment-field models used Adam optimization with batch size 4096, 80,000--90,000 steps, and learning rates between $9\times10^{-4}$ and $10^{-3}$ depending on $N$. All neural optimization results are reported for one fixed seed; seed variance was not evaluated.

\section*{S4. Cavity state and realizability}
The cavity state vector contains density, velocity components, temperature, heat-flux components, directional temperatures, stresses, selected third-order quantities, fourth-order quantities, and two closure diagnostics. The moment-field representation fits these channels directly; it does not guarantee positivity of an underlying VDF or enforce all thermodynamic identities by construction. This is why the main text treats the cavity model as a continuous wall-transport diagnostic, while the shock phase-space model is the representation that is positive by construction.

\section*{S5. Alt-text style figure descriptions}
Fig. 1: Six-panel comparison of DVM and Gaussian profiles for the $M=3$ normal shock, showing density, velocity, temperature, heat flux, normal stress, and third-order nonequilibrium moment. Fig. 2: The same comparison for the $M=5$ normal shock, with larger nonequilibrium discrepancies near the shock layer. Fig. 3: Cavity physical fields, comparing DVM, Gaussian reconstruction, and percent-range error for speed, vorticity, and temperature. Fig. 4: Cavity heat-flux and shear-stress fields, comparing DVM, Gaussian reconstruction, and percent-range error for $q_x$, $q_y$, and $\sigma_{xy}$.